\documentclass[aps,twocolumn,superscriptaddress,showpacs,floatfix]{revtex4-1}
\usepackage{graphicx}
\usepackage{times}
\usepackage{bm}
\usepackage{mathtools, nccmath}
\usepackage{xcolor}
\usepackage{amsmath}
\usepackage{amssymb}
\usepackage{euscript}   
\usepackage{verbatim}
\usepackage{fancyhdr}
\usepackage{wrapfig}
\usepackage{setspace}
\usepackage{xcolor}
\usepackage{amsfonts}
\usepackage{subfigure}
\usepackage{nicefrac}
\usepackage{hyperref}
\usepackage[mathlines]{lineno}
\usepackage{textcomp}
\usepackage{gensymb}
\usepackage[super]{nth}
\newcommand{\be}{\begin{equation}}
\newcommand{\ee}{\end{equation}}
\newcommand{\bea}{\begin{eqnarray}}
\newcommand{\eea}{\end{eqnarray}}

\begin{document}

\title{Helical superconducting edge modes from 
pseudo-Landau levels in graphene}

\author{Daniel Sabsovich}
    \affiliation{Raymond and Beverly Sackler School of  Physics and Astronomy, Tel Aviv University, Tel Aviv 69978, Israel}
\author{Marc W. Bockrath}
    \affiliation{Department of Physics, The Ohio State University, Columbus, OH 43210, USA}
\author{Kirill Shtengel}
    \affiliation{Department of Physics and Astronomy, University of California at Riverside, Riverside, California 92521, USA}     
\author{Eran Sela}
    \affiliation{Raymond and Beverly Sackler School of  Physics and Astronomy, Tel Aviv University, Tel Aviv 69978, Israel}

\begin{abstract}
We explore Andreev states at the interface of graphene and a superconductor for a uniform pseudo-magnetic field. 
 Near the zeroth-pseudo Landau level, we find a topological transition as a function of applied Zeeman field, at which a gapless helical mode appears. 
    This 1D mode is protected from backscattering as long as intervalley- and spin-flip scattering are suppressed.  We discuss a possible experimental platform to detect this gapless mode based on strained suspended membranes 
on a superconductor, in which dynamical strain causes charge pumping.
\end{abstract}


\maketitle

\draft

\vspace{2mm}

\section{Introduction}
Synthetic gauge fields resulting from strain 
~\cite{levy2010strain,Si2016strain,jiang2017visualizing,Deji2017review,nigge2019room} have been observed in numerous experiments on graphene samples~\cite{wallace1947Graphite,Novoselov2005,Zhang2005,Castro2009}. These gauge fields result from the movement of the Dirac points in momentum space, which results from the modification of the hopping amplitudes by strain~\cite{guineaenergy,vozmediano2010gauge}. This phenomenon is general for Dirac materials~\cite{pikulin2016chiral,grushin2016inhomogeneous,Ilan2019pseudo,gorbar2017chiral,arjona2018rotational,kamboj2019generation} and results in time-reversal symmetric pseudo-Landau levels (PLLs). 
In various geometries of graphene membranes, it is possible to engineer strain to yield uniform pseudo-magnetic fields~\cite{guinea2010generating,guineaenergy,zhu2015programmable,sela2020quantum}.

Since pseudo-magnetic fields act oppositely on  the two valleys of graphene, they 
may have a 
distinct interplay with superconductivity~\cite{covaci2011superconducting,Ghaemi2012,uchoa2013superconducting,Roy2014,AMORIM20161,lee2017fractional,Massarelli2017,liu2017quantum,Nica2018,Khanjani2018anomalous}. A usual superconducting order parameter involving pairing of time-reversal partners, is blind to such a field. Nevertheless, the microscopic theory, critical temperature and quasiparticle excitations are strongly affected by a pseudo-magnetic field.

\begin{figure}[t]
	\centering
	\includegraphics[scale=0.41]{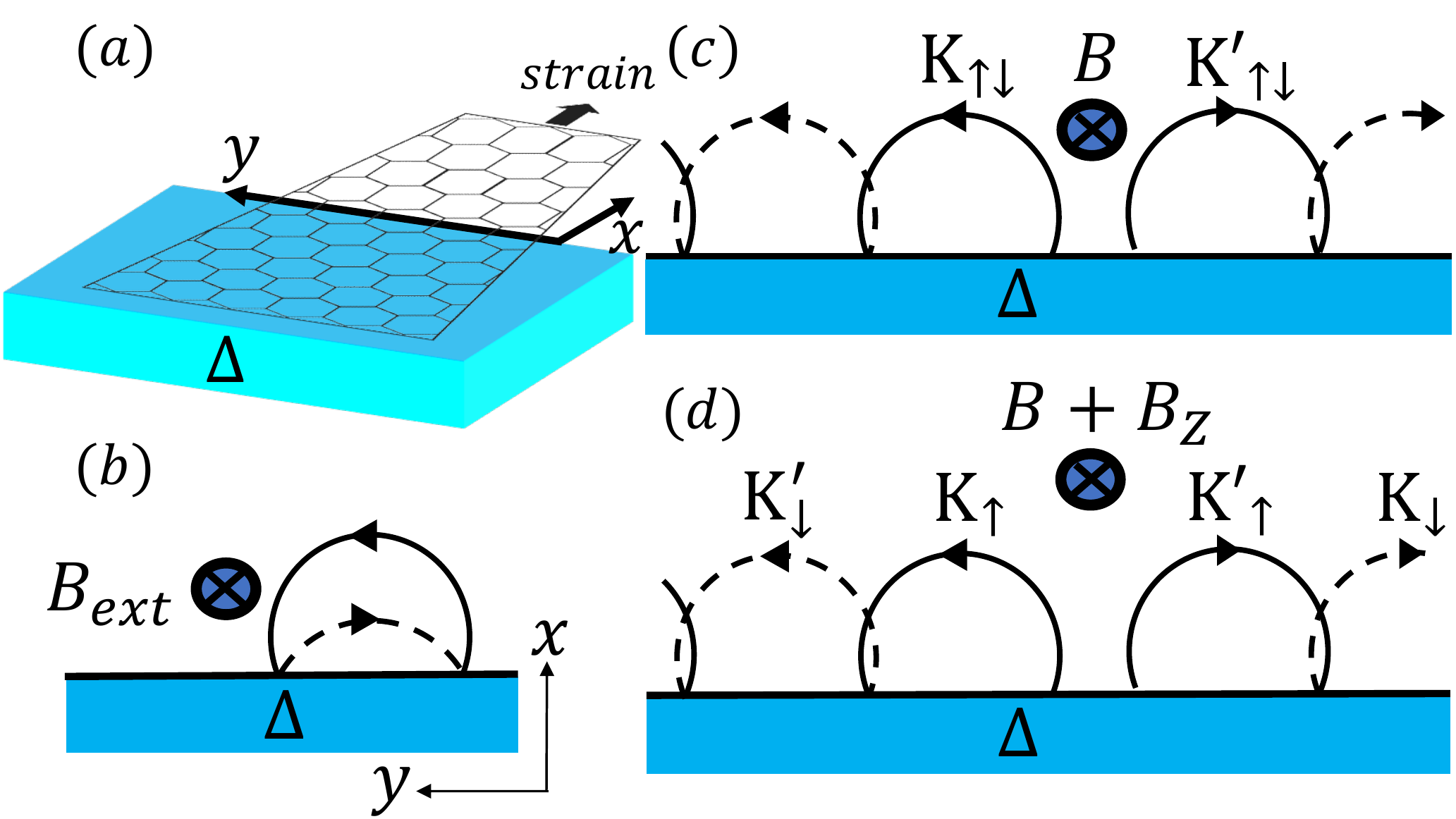}
	\caption{(a) Schematics of a graphene ribbon deposited on a superconductor $(\Delta)$ for $x<0$ and strained along $x$. The $(x=0)$ interface along $y$ is assumed to be infinite. (b) Andreev interface modes for graphene near the neutrality point, for a real magnetic field $B_{ext}$~\cite{akhmerov2007detection}, or (c) for a pseudo magnetic field $B$~\cite{gunawardana2015andreev}. (d) With an additional Zeeman  field $B_Z$ an Andreev helical mode is stabilized.
	}
	\label{fig1}\vspace{-0.2in}
\end{figure}

In this paper we explore 1D Andreev modes that can be stabilized on the interface between strained graphene and a superconductor~\cite{Andreev1964,beenakker2006specular}, see Fig.~\ref{fig1}(a). In the presence of a real magnetic field Akhmerov and Beenakker~\cite{akhmerov2007detection} found cyclotron orbits of Andreev reflected electrons and holes that are fully localized when the chemical potential lies at the Dirac point, 
see Fig.~\ref{fig1}(b). Physically, a specular-reflected hole~\cite{beenakker2006specular} proceeds along the mirror-reflected segment of the electronic cyclotron trajectory. 
In the case of a pseudo-magnetic field,  
Gunawardana and Uchoa~\cite{gunawardana2015andreev} found propagating states instead; see Fig.~\ref{fig1}(c). 
~By performing a mirror transformation to the hole trajectories and recalling that the pseudo-magnetic field has opposite signs for the two valleys, those solutions become snake states, propagating on interfaces at which a magnetic field changes sign. 

Here we study the interplay of the pseudo-magnetic field $B$ and a Zeeman magnetic field $B_Z$ on the graphene-superconductor interface, see Fig.~\ref{fig1}(a). We show that as $B_Z$ exceeds the chemical potential $\mu$ measured from the Dirac point, 
the  1D interface becomes gapless, hosting a helical 1D mode, see Fig.~\ref{fig1}(d). 
Helical modes consist of counter-propagating Dirac fermions carrying opposite spin, similar to those realized on the edge of a 2D topological insulator~\cite{konig2007quantum} or in the zeroth Landau level in graphene subjected to a Zeeman field~\cite{abanin2006spin,veyrat2020helical}.
On the interface between a superconductor and a strained graphene, the helical modes appearing due to the Zeeman field are gapless Andreev states protected from backscattering as long as (i) the disorder is smooth and does not result in intervalley scattering and (ii) spin flip is not possible. By contrast, in the case of a real magnetic field, superconducting pairing leads to the gapping of the helical edge modes, providing a platform for hosting Majorana zero modes at the boundaries of superconducting domains~\cite{san2015majorana}.

The paper is organized as follows. In Sec.~\ref{se:1Dmodes} we solve the BdG equations describing graphene within the  Dirac theory in the presence of a pairing potential step and pseudo-magnetic field. 
We explore in detail the low energy part undergoing a phase transition in Sec.~\ref{se:1DmodesloWE} by projecting down to the zeroth PLL and treating superconductivity, Zeeman field, and also spin-orbit coupling, as perturbations. 
In Sec.~\ref{se:pumping}
we discuss a Thouless pumping experiment, taking place as the flux associated with the pseudo-magnetic field varies in time. 
We conclude in Sec.~\ref{se:conc}.

\section{Interface modes}
\label{se:1Dmodes}
We consider a graphene sheet subjected to a uniform pseudo-magnetic field $B$, proximitized by a superconductor covering the $x<0$ region, see Fig.~\ref{fig1}(d). We 
write the Bogoliubov–de Gennes (BdG)  equation as
\be
\label{BdGmatrix}
\left[ {\begin{array}{cc}
	H-\mu & \Delta(x) \\
	\Delta(x)^* & \mu-T H T^{-1} \\
	\end{array} } \right] \Psi = \varepsilon \Psi,
	\ee
where $\Delta(x)=\Delta$ for $x<0$ and $\Delta(x)=0$ for $x>0$. $T$ here is the time-reversal operator and $\mu$ is the chemical potential.
The $8 \times 8$ Hamiltonian $H$  is decomposed as $H=H_0 + H_{Z}$, where $H_0$ describes the graphene and $H_{Z}$ is a Zeeman term.  Each piece can be written in terms of Pauli matrices $\tau,\sigma$ and $s$ acting in valley, pseudo-spin, and spin spaces, respectively. In the valley-symmetric notation
~\cite{akhmerov2007detection}, graphene in a pseudo-magnetic field is described by the Hamiltonian 
\be
H_0=v_F \sum_{i=x,y} \left( p_i \tau_0 \otimes \sigma_i+ A_i \tau_3 \otimes \sigma_i \right) \otimes s_0,
\ee
where $v_F$ is the Fermi velocity. Written explicitly in the valley space, it reads
\be
\label{MatrixEquation}
H_0=v_F \begin{pmatrix} (\mathbf{p}+\mathbf{A}) \cdot \sigma & 0\\ 0 & (\mathbf{p}-\mathbf{A}) \cdot \sigma \end{pmatrix}\otimes s_0.
\ee
Even in the presence of the pseudo-magnetic field, $H_0$ is time-reversal symmetric,  $H_0=TH_0T^{-1}$. Here $T=-i \tau_y \sigma_y s_y C$ where $C$ denotes complex conjugation, satisfying $T^2=-1$. On the other hand
$H_Z=B_Z \tau_0 \otimes \sigma_0 \otimes s_z$ describes a Zeeman field, which is time reversal odd, $H_Z=-TH_ZT^{-1}$.

In what follows, we use the Landau gauge $\mathbf{A}=B x\mathbf{\hat{y}}$. 
An eigenvector of the Hamiltonian in Eq.~(\ref{MatrixEquation}) in the $K$ valley can be written as $\phi_K e^{i p_y y}$ where $\phi_K(x)$ is a 2-spinor satisfying
\be
[  (-i \partial_x) \sigma_x + (p_y+x)\sigma_y] \phi_K= \varepsilon \phi_K.
\ee
We measure length in the units of magnetic length $\ell_B=\sqrt{{\hbar}/{eB}}$, momentum $p_y$ in the units of $\hbar/\ell_B$, and energy in the units of $\hbar v_F/\ell_B$.
The 2-spinor solutions are
\bea
\label{spinors}
\Phi_K^{(n)} =e^{-\frac{1}{2} \xi_+^2} \begin{pmatrix} -i \sqrt{2n} H_{n-1} (\xi_+) \\ \pm H_{n} (\xi_+) \end{pmatrix} _{\varepsilon=\pm  \sqrt{2n}}, \nonumber \\
\Phi_{K'}^{(n)} =e^{-\frac{1}{2} \xi_-^2} \begin{pmatrix} \pm H_{n} (\xi_-) \\ -i \sqrt{2n} H_{n-1} (\xi_-) \end{pmatrix} _{\varepsilon=\pm \sqrt{2n}},
\eea
where $\xi_\pm=x \pm p_y$ and $H_n(x)$ are Hermite polynomials. As detailed in  Appendix \ref{apendix0}, we proceed by using these solutions for the normal region $x>0$ and imposing a 
boundary condition describing Andreev reflection~\cite{titov2006josephson,akhmerov2007detection}, which is valid when the superconducting coherence length is smaller than $\ell_B$.


\begin{figure}[t]
\includegraphics[width=0.9\columnwidth, keepaspectratio=true]{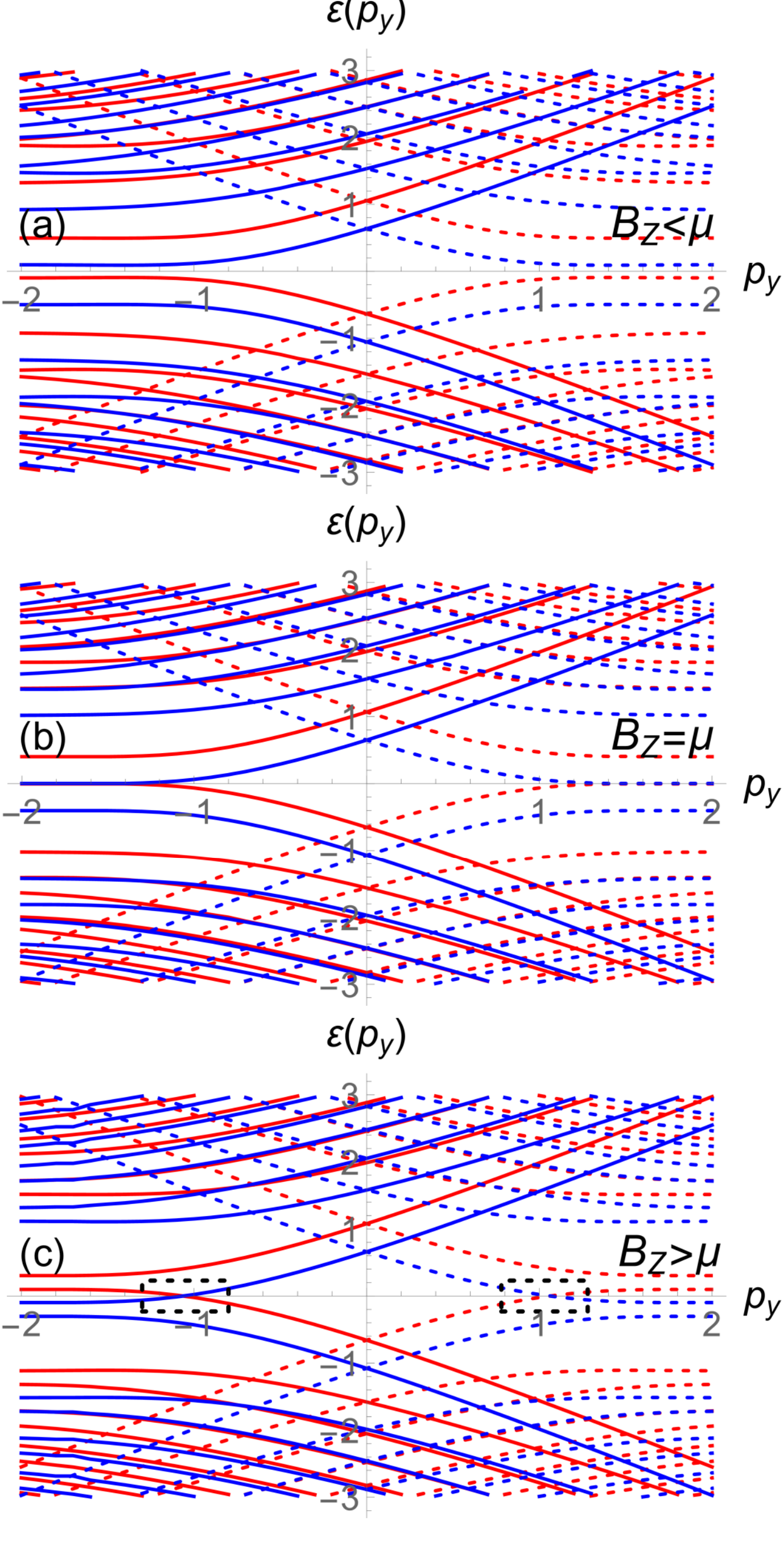}\hfill
\caption{Dispersion relation $\varepsilon(p_y)$ of Andreev states at a 1D interface between proximitized graphene and normal graphene with PLLs, obtained by solving Eq.~(\ref{AB1}) for $\Delta=10$, $B_Z=0.2$. We use red/blue to describe opposite spins, and full/dashed lines to describe opposite valleys. Panel (a) shows a gapped spectrum at  $\mu=0.3$; (b) the gap closes at $\mu=0.2$; (c) gapless helical interface modes are marked by dashed rectangles for $\mu=0.1$.}
\label{fig:res}\vspace{-0.1in}
\end{figure}

\begin{figure}[t]
\includegraphics[width=0.9\columnwidth, keepaspectratio=true]{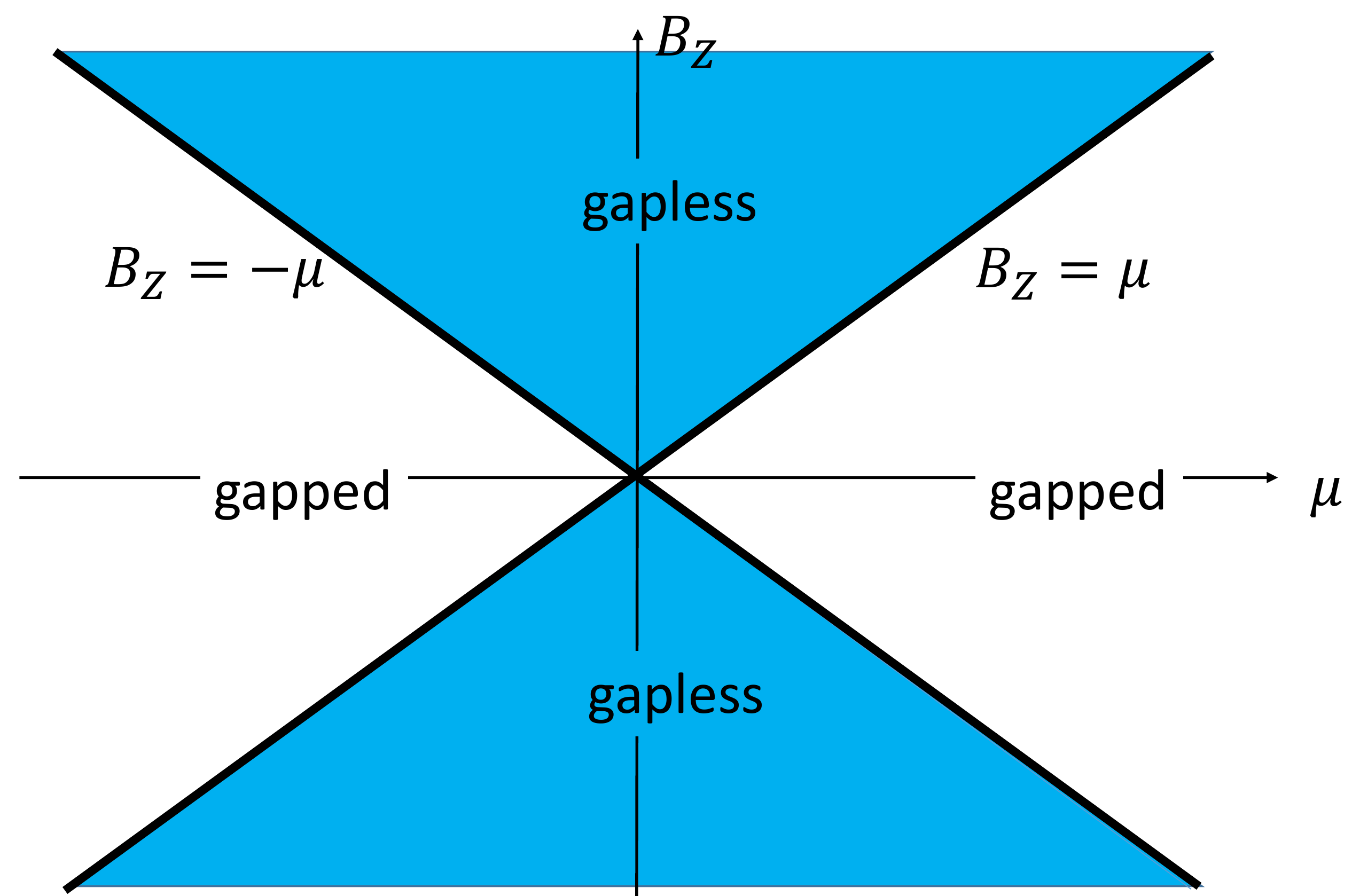}\hfill
\caption{Phase diagram of our superconductor-normal (SN) interface for $B_Z,\mu$ smaller than the PLL spacing.}
\label{fig:phasediagram}\vspace{-0.1in}
\end{figure}

The  dispersion relation $\varepsilon(p_y)$ is plotted in Fig.~\ref{fig:res}. 
The momentum along the interface, $p_y$, determines the position along the $x$-axis; for $K$($K'$) the electron wave function is localized near $x = -  p_y$($x=p_y$). Solid lines represent valley-$K$ electron states, which for negative momentum approach the $\pm \sqrt{2n}$ values of the PLLs for $B_Z=\mu=0$ (not shown), and correspond to states localized in the normal region. For increasing $p_y$, the modes acquire a dispersion due to the pairing potential $\Delta(x)$ inside the superconducting region. The opposite $p_y$-$x$ relation applies for $K'$ states (dashed lines).


This model has two phases in the $\mu-B_Z$ plane. For $|\mu|<|B_Z|$, we have gapless 1D interface modes, as seen in Fig.~\ref{fig:res}(c). By contrast, for $|\mu| > |B_Z|$, the interface is gapped, as seen in Fig.~\ref{fig:res}(a). 
At the transition $|\mu|=|B_Z|$ [see Fig.~\ref{fig:res}(b)] the PLL excitation gap closes at the normal side. The corresponding phase diagram is shown in Fig.~\ref{fig:phasediagram} for the regime where $\mu$ and $B_Z$ are small compared to the PLL separation $\hbar v_F/\ell_B$.

In Appendix \ref{apendix2} we demonstrate the emergence of gapless 1D interface modes using a tight-binding model, reproducing the same low energy physics even when additional lattice effects are present, such as zigzag edge states.  In Appendix \ref{apendix0} we discuss what happens when the Zeeman field $B_Z$ exceeds the PLL gap. Below we  focus on the low energy part by restricting the theory to the zeroth PLL.  

\section{Projection to the zeroth PLL}
\label{se:1DmodesloWE}
We proceed by projecting the Hamiltonian to the zeroth PLL. This approach is valid for $B_Z , \mu , \Delta \ll \hbar v_F/\ell_B$. The normalized zero energy solutions of $H_0$ in Eq.~(\ref{spinors}) are
\be
\label{zenespinor}
\Phi_K^{(n=0)} =\frac{1}{\pi^{1/4}}e^{-\frac{1}{2} \xi_+^2} \begin{pmatrix} 0 \\ 1 \end{pmatrix} ,\quad 
\Phi_{K'}^{(n=0)} =\frac{1}{\pi^{1/4}} e^{-\frac{1}{2} \xi_-^2} \begin{pmatrix} 1 \\ 0 \end{pmatrix} .
\ee
We use the valley-symmetric notation in which the two sublattices are interchanged for valley $K'$.
For a given $p_y$ the dimension of the degenerate zero energy space of the BdG matrix Eq.~(\ref{BdGmatrix}) is 8, which includes an additional duplication due to particles and holes.
We introduce a basis 
$|\tau^z,s^z,\eta^z \rangle$ ($\tau^z,s^z,\eta^z=\pm$) denoting the valley, spin, and particle-hole spaces, respectively. Equivalently, in Eq.~(\ref{BdGmatrix}) we can use the Bogoliubov spinor 
\be
\label{8spinorA}
\Psi = 
\left(\psi_{K \uparrow},\psi_{K \downarrow},\psi_{K' \uparrow},\psi_{K' \downarrow},\psi^\dagger_{K' \downarrow},-\psi^\dagger_{K' \uparrow},\psi^\dagger_{K \downarrow},-\psi^\dagger_{K \uparrow}\right)
\ee
whose eight components are defined in the basis $|\tau^z,s^z,\eta^z \rangle$ as
\bea
\label{8spinorB}
&&|+++\rangle,|+-+\rangle,|-++\rangle,|--+\rangle, \nonumber \\ &&|++-\rangle,|+--\rangle,|-+-\rangle,|---\rangle \equiv |1\rangle , \dots |8 \rangle.
\eea


\begin{figure}[t]
\includegraphics[width=0.89\columnwidth, keepaspectratio=true]{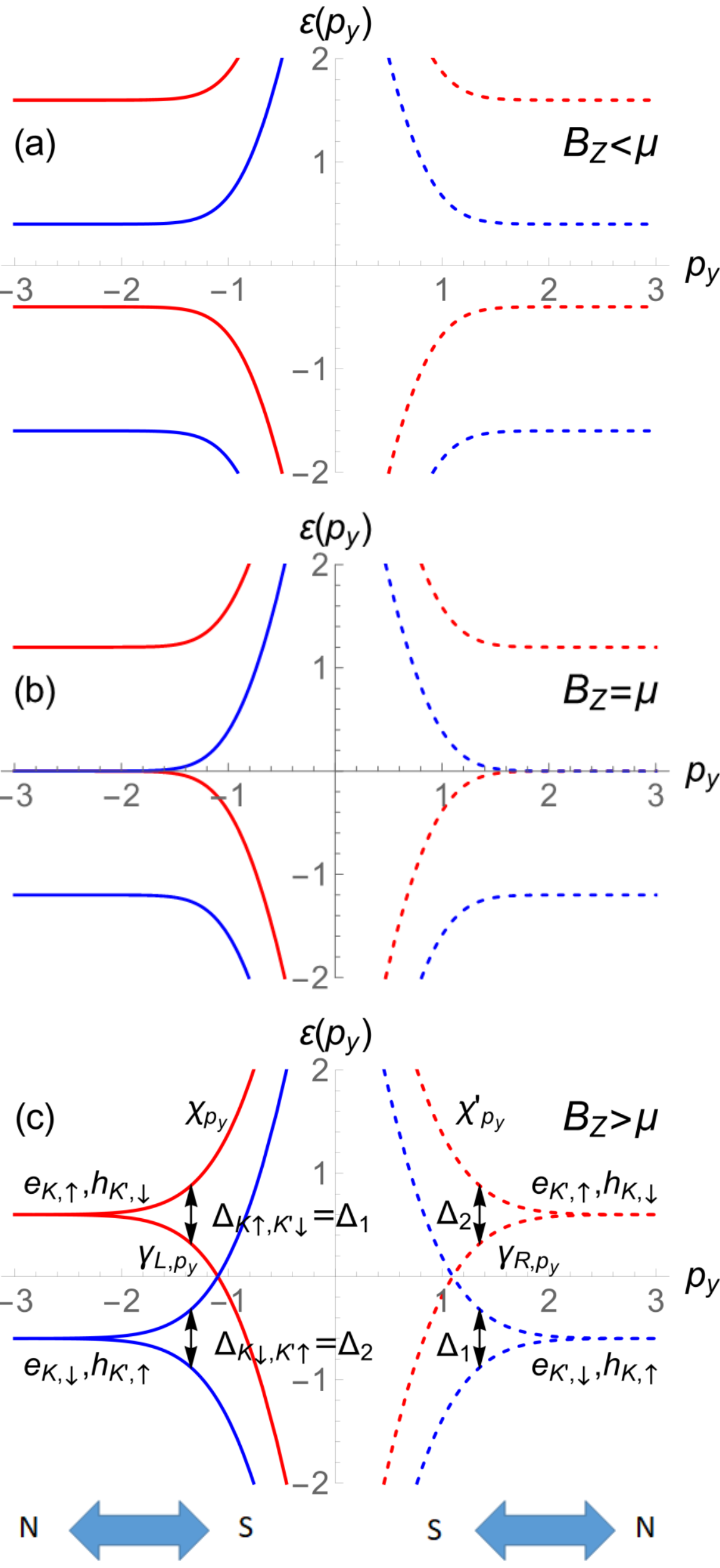}\hfill
\caption{Dispersion relation of our SN interface (as in Fig.~3) focusing on the zeroth PLL using Eq.~(\ref{eq:lowH0LL}), 
setting $\Delta=10$ and $B_Z=0.6$. In (a) $\mu=1$ and we have a gapped interface. As we decrease $\mu$ the gap decreases and eventually closes in (b) at $\mu=0.6$. The condition $|\mu|<|B_Z|$ leads to a gapless helical Dirac fermion as seen in (c) where $\mu=0$. At the bottom we illustrate the 
$x-p_y$ relationship.}
\label{fig:reslowE}\vspace{-0.4in}
\end{figure}

We now project the various terms of the BdG Hamiltonian in Eq.~(\ref{BdGmatrix}). Consider the pairing part of the Hamiltonian first: 
$$\mathcal{H}_\Delta=\begin{pmatrix}0 & \Delta(x)\\
\Delta^{*}(x) & 0
\end{pmatrix}.$$
It has matrix elements
\bea
\label{2spinor}
\langle + s^z \mp | \mathcal{H}_\Delta |+ s^z \pm \rangle &=& \Delta \int_{-\infty}^0 \frac{1}{\sqrt{\pi}}e^{-(x+p_y)^2} dx \equiv  \Delta ~ F(p_y),\nonumber \\ 
\langle - s^z \mp | \mathcal{H}_\Delta |- s^z \pm \rangle &=&  \Delta~ F(-p_y).
\eea
The function $F(p_y)$, describing the matrix element of the pairing potential in a state with momentum $p_y$, behaves as $F(p_y) \to 1$ for $p_y \to \infty$ and $F(p_y) \cong \frac{e^{-p_y^2}}{2 \sqrt{\pi}|p_y|}$  for $p_y \to -\infty$. Physically, for positive $p_y$ the support of wave functions in  Eq.~(\ref{zenespinor}) corresponding to valley $K$ is well within the superconductor, while for negative $p_y$ it is mainly in the normal region. In the latter case the wave functions are affected by the pairing potential only through their exponential tail. Hence,
\begin{multline}
\mathcal{H}_\Delta(p_y)=\frac{\Delta}{2} \left[F(p_y)(\tau_0+\tau_z)\right.\\
\left. + F(-p_y)(\tau_0-\tau_z)\right] \otimes s_0 \otimes \eta_x.
\end{multline}
Similarly, the Zeeman term takes the form
\be
\label{ZeemanHamiltonain}
\mathcal{H}_Z(p_y)= B_Z \tau_0 \otimes s_z \otimes \eta_0.
\ee
The eigenvalues of the projected zeroth-PLL Hamiltonian
\be
\label{eq:lowH0LL}
\mathcal{H}_{0LL}(p_y) =\mathcal{H}_Z(p_y)+\mathcal{H}_\Delta (p_y)-\mu \eta_z ,
\ee
are plotted in Fig.~\ref{fig:reslowE} for various values of $B_Z$ and $\mu$, reproducing the low energy sector seen in Fig.~\ref{fig:res}. 

Let us focus on the four solutions at large negative $p_y$, which are basically unaffected by $\Delta$. They originate from electrons in valley $K$ or holes of valley $K'$ (not dashed). 
 Their energies are
\bea
\label{Eorder}
&&\varepsilon_1(e_{K\uparrow})=B_Z-\mu - \lambda_I, \nonumber \\
&&\varepsilon_2 (e_{K\downarrow})=-B_Z-\mu + \lambda_I,\nonumber \\
&&\varepsilon_5 (h_{K'\downarrow})=B_Z+\mu+ \lambda_I,\nonumber \\
&&\varepsilon_6(h_{K' \uparrow})= -B_Z+\mu- \lambda_I .
\eea
In Sec.~\ref{se:SI} we will introduce the spin-orbit coupling $\lambda_I$  and discuss its role; for now we set $\lambda_I=0$. Here the subscripts refer to components of the Bogoliubov spinor $\Psi$, see Eqs.~(\ref{8spinorA}) and (\ref{8spinorB}).
We can see in Fig.~4 that $\Delta$ separately couples modes 1 and 5 ($K\uparrow$ and $K' \downarrow$) as marked by red curves which diverge as $p_y$ increases, and similarly 2 and 6 ($K\downarrow$ and $K' \uparrow$) shown as diverging blue curves. Let us denote the corresponding level separations by $\Delta_{1,2}$, respectively.
As can be seen in Fig.~\ref{fig:reslowE}(c), in the topological phase $|B_Z|>\mu$  this level repulsion leads to zero energy edge states.


Consider the left-moving gapless quasiparticles denoted $\gamma_L$ originating from the $p_y<0$ region due to the $\Delta_1$ sector in Fig.~\ref{fig:reslowE}(c). Along with the gapped partner denoted $\chi$, this pair [red~solid curves in Fig.~\ref{fig:reslowE}(c)] can be expressed as a combination of $\psi_{K \uparrow ,p}$ and $\psi^\dagger_{K' \downarrow ,-p}$ in the form
\bea
\label{eq:gammaL}
&&{\rm{gapless}}: \gamma_{L,p_y}=u_{p_y} \psi_{K \uparrow ,p_y}+v_{p_y} \psi^\dagger_{K' \downarrow ,-p_y},\nonumber \\
&&{\rm{gapped}}: \chi_{p_y}=-v_{p_y} \psi_{K \uparrow ,p_y}+u_{p_y} \psi^\dagger_{K' \downarrow ,-p_y}.
\eea
Similarly, consider the red dashed pair of curves in Fig.~\ref{fig:reslowE}(c) forming the $\Delta_2$ sector. They lead to a right moving gapless mode $\gamma_R$, and a gapped mode denoted $\chi'$
\bea
\label{eq:gammaR}
&&{\rm{gapless}}: \gamma_{R,p_y}=u'_{p_y} \psi_{K' \uparrow ,p_y}+v'_{p_y} \psi^\dagger_{K \downarrow ,-p_y},\nonumber \\
&&{\rm{gapped}}: \chi'_{p_y}=-v'_{p_y} \psi_{K' \uparrow ,p_y}+u'_{p_y} \psi^\dagger_{K \downarrow ,-p_y}.
\eea
These 1D modes are Dirac- rather than Majorana-fermions. Namely $\gamma_L \ne \gamma_L^\dagger$ and $\gamma_R \ne \gamma_R^\dagger$. 
Considering for example 
$\gamma_R$ in the $p_y>0$ region (red-dashed), given in Eq.~(\ref{eq:gammaR}), and applying hermitian conjugation, one obtains the right moving gapless mode in the $p_y<0$ region (blue-solid). Thus we have both particle and hole excitations, formed out of the quasiparticle operators, allowing to define Dirac fermions. Linearizing the spectrum near $\varepsilon=0$, the low energy Hamiltonian is 
\begin{multline}
H_{{\rm{Dirac}}}=v_0 \sum_{p_y} \left[(p_y - p_0) \gamma^\dagger_{R,p_y}\gamma_{R,p_y}
\right.\\
\left.- (p_y + p_0) \gamma^\dagger_{L,p_y}\gamma_{L,p_y} \right].\label{eq:DiracHamil}
\end{multline}
It is valid for energies low compared to ${\rm{min}} \{\varepsilon_1,\varepsilon_2,\varepsilon_5,\varepsilon_6 \}$, see Eq.~(\ref{Eorder}).

The resulting counter-propagating Dirac modes are similar to the edge states in the quantum-spin Hall effect. While we consider non-interacting electrons, interactions as well as disorder on the edge can be treated as it is done in the context of the quantum spin Hall effect~\cite{xu2006stability,wu2006helical,sela2011majorana}.

\begin{figure}[t]
\includegraphics[width=0.85\columnwidth, keepaspectratio=true]{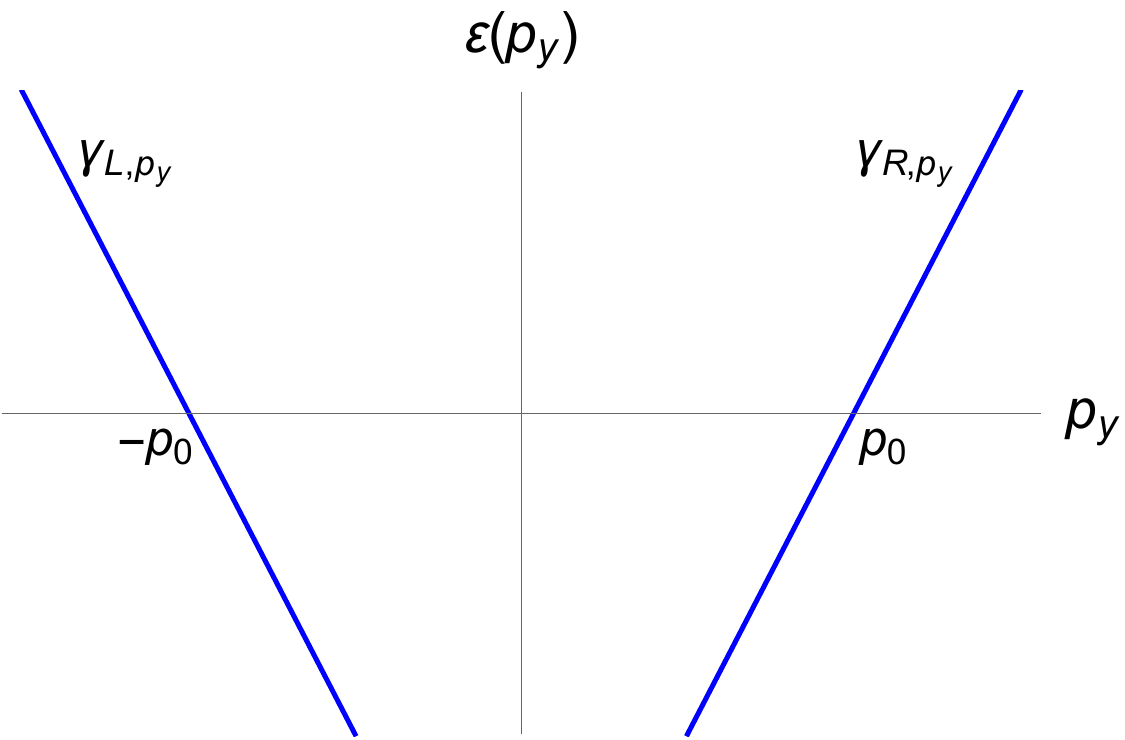}\hfill
\caption{Low energy spectrum describing the helical Dirac mode. the right and left moving excitations are described in Eqs.~(\ref{eq:gammaL}-\ref{eq:gammaR}), and the Dirac Hamiltonian is given in Eq.~(\ref{eq:DiracHamil}).}
\label{fig:HelicalModes}\vspace{-0.0
in}
\end{figure}

\subsection{$\mathbb{Z}_2\times \mathbb{Z}_2$ symmetry}

Usually proximity-induced superconductivity 
$\sum_i \Delta c^\dagger_{i \uparrow} c^\dagger_{i \downarrow}$ breaks U(1) charge conservation down to $\mathbb{Z}_2$, the parity conservation:
\be
P=e^{i \pi \sum_{j,\sigma} c^\dagger_{j,\sigma} c_{j,\sigma}}.
\ee
At energy sufficiently low compared to the band width of graphene, the pairing Hamiltonian becomes $\int dr [\Delta_1 \psi^\dagger_{K \uparrow} \psi^\dagger_{K' \downarrow}+\Delta_2  \psi^\dagger_{K' \uparrow} \psi^\dagger_{K \downarrow} +h.c]$, with the two terms accounting for the two valleys at opposite momenta. While $\Delta_1 = \Delta_2=\Delta$, this form emphasizes that we  have two conserved parities,
\bea
\label{z2z2}
P_1&=&e^{i \pi  \int d^2 r (\psi^\dagger_{K \uparrow}\psi_{K \uparrow} + \psi^\dagger_{K' \downarrow}\psi_{K' \downarrow} )},\nonumber \\
P_2&=&e^{i \pi  \int d^2r (\psi^\dagger_{K \downarrow}\psi_{K \downarrow} + \psi^\dagger_{K' \uparrow}\psi_{K' \uparrow} )}.
\eea
This $\mathbb{Z}_2\times \mathbb{Z}_2$ parity symmetry applies as long as the following two processes are suppressed: inter-valley scattering, $H_{\text{i-v}} =  V_{\text{iv}} \sum_{\sigma,p_y}  \psi^\dagger_{K \sigma,p_y} \psi_{K' \sigma,p_y}+\text{h.c.}$, and the spin-flip scattering, $H_\text{s-f} =V_{\text{sf}} \sum_{p_y}  \psi^\dagger_{K \uparrow,p_y} \psi_{K \downarrow,p_y}+\psi^\dagger_{K' \uparrow,p_y} \psi_{K' \downarrow,p_y} +\text{h.c.}$. Both processes break the separate conservation of $P_1$ and $P_2$ down to a single $\mathbb{Z}_2$ symmetry corresponding to the conservation of $P_1 \cdot P_2$. Expressed in terms of the low energy quasiparticle excitations, $H_{\text{i-v}}$ becomes $H_\text{i-v}  \sim V_\text{iv} \sum_{p_y} \gamma^\dagger_{L,p_y} \gamma_{R,p_y}+\text{h.c.}$.
A spatially uniform inter-valley scattering term does not open a gap since generally scattering between the two Dirac modes requires momentum transfer, see Fig~\ref{fig:HelicalModes}. However, sufficiently short-range disorder can enable such momentum transfer, and hence backscattering. 
In the quasiparticle language $H_{\text{s-f}}$ becomes $H_{\text{s-f}} \sim V_\text{sf} \sum_{p_y} \gamma^\dagger_{L,p_y} \gamma^\dagger_{R,-p_y} + \text{h.c.}$. Naively, one would expect that the level crossing at $\varepsilon=0$ would  become avoided for the Zeeman magnetic field tilted away from the $z$ axis. However, this intuition is incorrect and the gapless mode persists irrespective of the direction of the field as long as a $U(1)$ conservation of the spin along some direction persists. In that case one can redefine a corresponding conserved $\mathbb{Z}_2\times \mathbb{Z}_2$. However, as discussed in the next subsection, in the presence of spin-orbit coupling, tilting the direction of the Zeeman field does open a gap due to the quasiparticle backscattering.

In general zero energy solutions to the BdG equation 
reflect a degeneracy between ground states with different parity. In our model, for $|B_Z|>|\mu|$ we have $\varepsilon \to 0$ quasi-particle solutions which change either $P_1$ ($\gamma_L$) or $P_2$ ($\gamma_R$), allowing us to identify four ground states. 

It is interesting to note that the association of left movers with  $\Delta_{1}=\Delta_{K \uparrow,K' \downarrow}$  and of the right movers with $\Delta_{2}=\Delta_{K' \uparrow,K \downarrow}$ switches upon switching the sign of the Zeeman field.

\subsection{Spin orbit coupling}
\label{se:SI}
Intrinsic spin orbit coupling (SOC) in graphene, which in the standard notation~\cite{kane2005quantum} takes the form $\lambda_I \tau_z \otimes \sigma_z \otimes s_z$, becomes in our valley-symmetric notation 
\be
\label{SOCHamiltonain}
H_\text{SO}=\lambda_I \tau_0 \otimes \sigma_z \otimes s_z.
\ee
Projecting this time-reversal symmetric term to the zeroth PLL subspace, its contribution to the BdG matrix Eq.~(\ref{BdGmatrix}), using Eq.~(\ref{zenespinor}), becomes
\be
\label{eq:SOC0LL}
\mathcal{H}_\text{SO}=-\lambda_I \tau_z \otimes s_z \otimes \eta_z.
\ee

\begin{figure}[t]
\includegraphics[width=0.9\columnwidth, keepaspectratio=true]{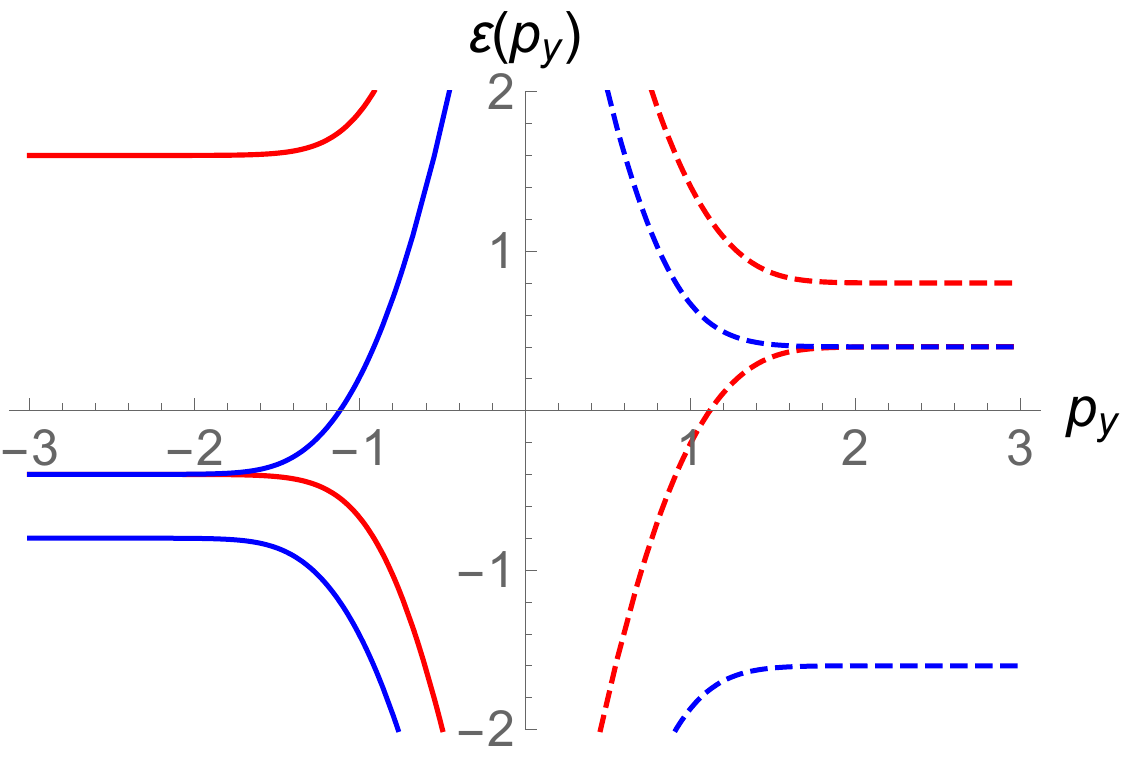}\hfill
\caption{Low energy dispersion from Eq.~(\ref{eq:lowH0LL}) 
with the addition of SOC Eq.~(\ref{eq:SOC0LL}). Here $\Delta=10$, $B_Z=\lambda_{I}=0.6$ and $\mu=0.4$.}
\label{fig:Pseudo_SOC_Dispersion}\vspace{-0.1
in}
\end{figure}

While the phase diagram in Fig.~\ref{fig:phasediagram} reflects a phase transition occurring in both $\mathbb{Z}_2$ sectors, in the presence of SOC one can induce a phase transition separately in each sector. One can generalize the arguments about the level repulsion due to $\Delta_1$ or $\Delta_2$ for the presence of SOC. In this case the energies of the four states in the normal region are given in Eq.~\ref{Eorder} with $\lambda_I$ included. The condition for the topological phase in the $\Delta_1$ sector ($\Delta_2$ sector) is that $\varepsilon_1$ and $\varepsilon_5$ ($\varepsilon_2$ and $\varepsilon_6$) have opposite signs. 
One can achieve these conditions independently. An example is shown in Fig.~\ref{fig:Pseudo_SOC_Dispersion}, where only the $\Delta_2$ sector is topological, with two right-moving branches, composing one right moving chiral Dirac fermion, and no left movers.  This exemplifies a quantum Hall state stabilized by a Zeeman field and SOC. In this case the edge states are chiral and hence protected against either spin-flip or intervalley scattering.


\subsection{Other pairing channels}
We remark that we have considered s-wave pairing in Eq.~(\ref{BdGmatrix}) but our model allows to consider more general pairings. In our $\mathbb{Z}_2\times \mathbb{Z}_2$ decomposition, s-wave pairing is reflected in equal pairings in each $\mathbb{Z}_2$ sector, $\Delta_1 = \Delta_2$, where we recall that $\Delta_{1}=\Delta_{K \uparrow,K' \downarrow}$  and  $\Delta_{2}=\Delta_{K' \uparrow,K \downarrow}$. However, one can consider any combination of singlet and triplet pairings, $\Delta_1 = \Delta_s +\Delta_t$, $\Delta_2 = \Delta_s -\Delta_t$. 
This provides an additional knob for tuning the topological transition in each $\mathbb{Z}_2$ sector.
Here we remark that close to the extreme case of equal superposition of singlet and triplet pairing, implying $\Delta_2=0$ (or $\Delta_1=0$), similar to the case of strong spin-orbit coupling in Fig.~\ref{fig:Pseudo_SOC_Dispersion}, we can have a situation where only the $\Delta_1$ (or $\Delta_2$) sector is topological. 

\subsection{Symmetry classification}
\label{se:AZ}
Like any weakly interacting fermionic system, our model can be classified according to the Altland--Zirnbauer symmetry classes. The superconducting system considered here has neither time-reversal nor spin-rotation symmetry, hence it belongs to class $D$~\cite{chiu2016classification} (see also table I in  Ref.~\onlinecite{kennedy2016bott}). Since our system has conserved $S^z$, implying a superconductor with $\mathbb{Z}_2\times \mathbb{Z}_2$ parity symmetry, its symmetry classification corresponds to two sectors each of which is in class $D$.

Class $D$ superconductors in two dimensions are characterized by a topological $\mathbb{Z}$ index, counting the number of chiral Majorana fermions on the edge with vacuum. The 1D interface under consideration here, on the other hand, is an interface with a normal system (non-superconducting symmetry class $A$). Nevertheless, as we discuss in Appendix~\ref{apendix2_1}, the interface with vacuum has indeed edge states when the Zeeman field exceeds the pairing gap. It is interesting that the two parity sectors generically can have unrelated $\mathbb{Z}$ indices whereas a helical-like state is characterized by two opposite indices. The occurrence of Dirac rather than Majorana fermions implies that in our system the $\mathbb{Z}$ index is restricted even integers.

\section{Pumping}
\label{se:pumping}


Consider a circular geometry 
with the superconductor covering the region $r >R$.
~Imagine that the pseudo-magnetic field is uniform in space 
and changes in time slowly. 
As we explain next, in this dynamical process charge is pumped into or out of the normal region, through the NS interface. 

First consider the region $\Delta =0$ at $r<R$. 
$dB/dt$ generates an azimuthal pseudo-electric field $\vec{E}=-\frac{d\vec{A}}{dt}$; Together with the pseudo-magnetic field we have a drift velocity $\vec{v}_d =\frac{\vec{E} \times \vec{B}}{|B|^2}$ along the radial direction, see Fig.~\ref{fg:flow}(a). The drift velocity has the same sign for the two valleys
~\cite{sela2020quantum}. The resulting charge current into a region of radius $r$ is
\be
\frac{dQ(r)}{dt} = \oint_r \vec{j}(r) \cdot d \vec{\ell}=2 \pi r n_e e \frac{E(r)}{B},
\ee
where we used $j=n_e e v_d$ for the inward charge current density. Assuming that the density, $n_e$, 
is determined by the filling factor $\nu$ as $n_e=\frac{B}{\Phi_0} \nu$, we have
\be
\frac{dQ(r)}{dt} = n_e e \frac{\pi r^2 \dot{B}}{B}=  e \frac{ \dot{\Phi}}{\Phi_0} \nu.
\ee
Thus the change in the number of particles coincides with $\nu$ times the change in the number of flux quanta of the pseudo-magnetic field. 

Equivalently, consider the adiabatic evolution of the many-body wave function generated by $\dot{B}$. 
~In the radial geometry $p_y$ corresponds to angular momentum and is quantized, together with the radius of the ring-orbitals, as \be
p_j = \frac{\hbar}{\ell_B} \sqrt{j}, ~~~~ r_j = \ell_B \sqrt{j},~~~j \in \mathbb{N}.
\ee
The values of $p_j$ and $r_j$ change upon increasing $B$ in such a way that the $r_j$'s become denser, see Fig.~\ref{fg:flow}(b). The particle density remains locked to the instantaneous value of $B$ according to $n_e=\frac{B}{\phi_0} \nu$.

\subsection{Pumping through the interface}
\begin{figure}[t]
\includegraphics[width=0.9\columnwidth, keepaspectratio=true]{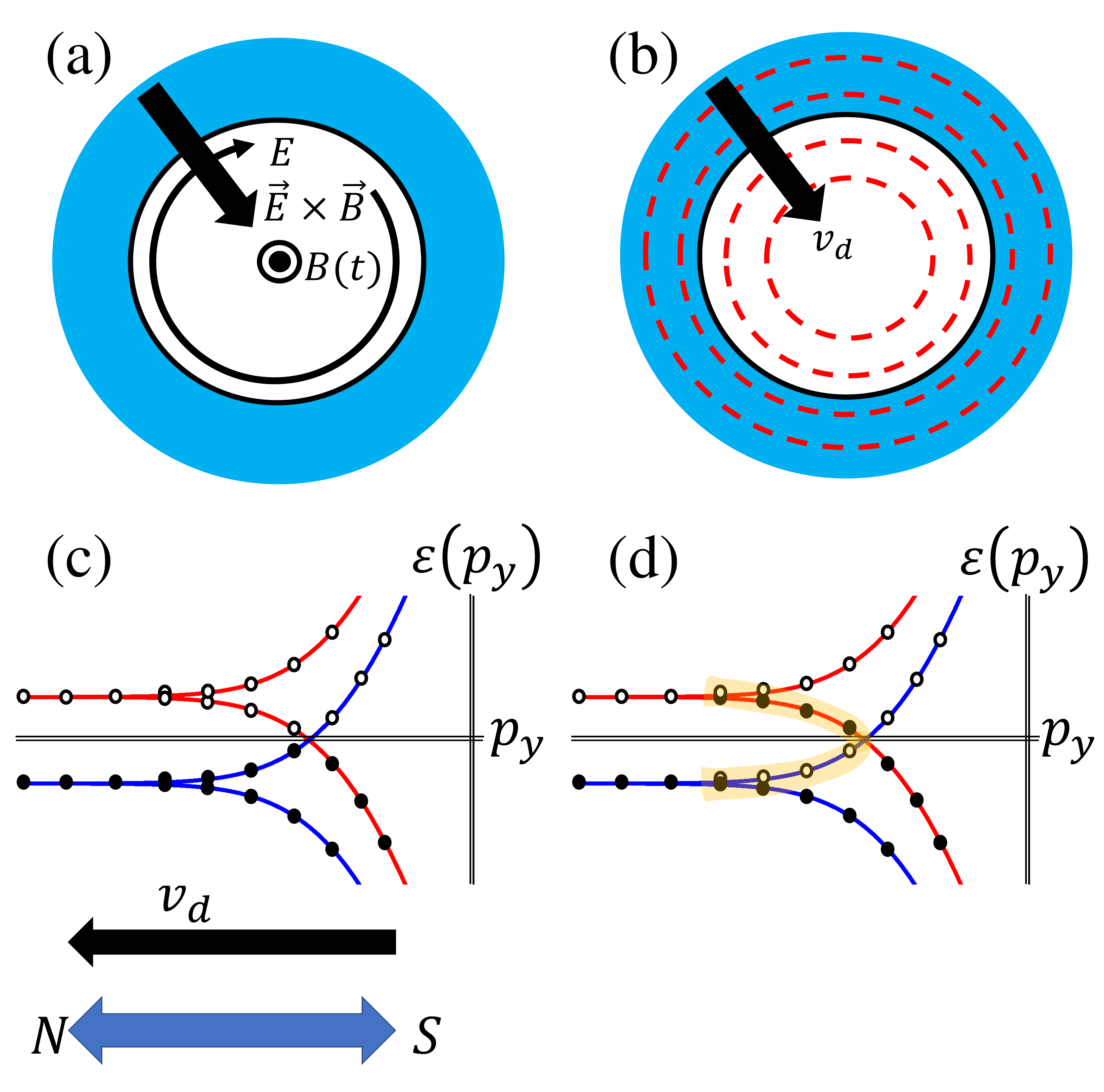}\hfill
\caption{Charge pumping: (a) Circular NS interface with a time dependent pseudomagnetic field $\dot{B}$ creating a pseudo-electric field $E$. 
The crossed $E$ and $B$ pseudo-fields result~\cite{sela2020quantum} in a radial charge current. (b) Ring-like PLL solutions adiabatically shrink as $B$ increases. 
(c) Quantized momenta 
and their occupation in the instantaneous ground state. These bands correspond to the left side $(p_y<0)$ of Fig.~5(c). The black arrow denotes the drift in $p_y$ as $B$ increases.
~In the $p_y>0$ region the drift is in the opposite direction in momentum space, and in the same spatial direction, not shown.  (d) Excited state resulting from pumping. Marked are excited single-quasiparticle states within the gapless helical branch. 
}
\label{fg:flow}\vspace{-0.1in}
\end{figure}

Now we discuss the role of the SN interface $r \sim R$. 
Does the many-body state resulting from the motion in $p_y$-space coincide with the instantaneous ground state? The answer is positive if the BdG spectrum of the interface is gapped, as in the trivial phase $|\mu|>|B_Z|$. 
However, the answer is negative for $|B_Z|>|\mu|$. In this case, Fig.~\ref{fg:flow}(c) represents schematically the occupation of BdG states near the left crossing in Fig.~\ref{fig:reslowE}(c) before pumping. We can see that occupied states form only subsets of BdG bands. As the spectral flow occurs, from the superconductor to the normal region, the $p_j$'s move to the left. The resulting many-body state in Fig.~\ref{fg:flow}(d) is an excited state.

Thus, the gapless  bands at the interface influence pumping by creating excitations. While we do not treat this explicitly, relaxation will eventually occur and create dissipation. On the other hand no dissipation is expected in the gapped regime $|B_Z|<|\mu|$.

\subsection{Possible experimental realization}
Finally, we mention a possible realization of this system.
On a practical level, our setup requires systems with controlled pseudo-magnetic fields.
Designs of  graphene membranes with programmable strain in order to engineer nearly uniform pseudo-magnetic fields over macroscopic distances have been proposed~\cite{guinea2010generating,guineaenergy,zhu2015programmable,sela2020quantum}. One such platform consists of a graphene flake suspended on top of a triangular aperture,  with the sides normal to the  $\langle 100 \rangle$ crystallographic axes of the graphene membrane. In this system a few-Tesla pseudo-field can be generated over a micrometer scale by electrostatically pulling the membrane towards a gate~\cite{guineaenergy}. 
This platform can then allow to induce pairing correlations on PLLs using a superconducting substrate. 

In the same system with the pseudo-magnetic field controlled by a gate, one can pump electrons in and out of the membrane simply by oscillating the gate potential, while the nanomechanical quality factor and frequency of the membrane could in principle be detected using approaches such as those described in Ref.~\cite{weber_coupling_2014}. The charge pumping itself can be detected via a shift of the mechanical resonance frequency. An explicit treatment of pumping in such a device remains a subject a future study. 

Another possible detection scheme is based on a rectification of the AC pumped current. One may create a current path from one side of the membrane to the other~\cite{low2012electron}. A rectified DC current could be created e.g. using an additional AC control of the relative resistance between the membrane and the two contacts.

\section{Conclusions}
\label{se:conc}
We studied a 1D interface between graphene with a strain-induced pseudo-magnetic field and a superconductor. Adding a Zeeman magnetic field, we identified a  phase that supports helical  edge modes. They are protected by
a $\mathbb{Z}_2 \times \mathbb{Z}_2$ symmetry, reflecting the separate conservation of either $(K,\uparrow)$ and $(K',\downarrow)$ electrons, or of $(K',\uparrow)$ and $(K,\downarrow)$ electrons. This emergent symmetry of proximitized graphene becomes exact when intervalley and spin-flip scatterings are suppressed. SOC allows to eliminate either the right-going or the left-going edge modes, resulting in a chiral edge mode.


Recently there have been numerous experiments exploring the interface of graphene with superconductors, in the presence of real magnetic fields \cite{Amet2016supercurrent,shalom2016quantum,lee2017inducing,draelos2018investigation,Zhu2018Supercurrent,lee2018proximity,huang2020interference,gul2020induced}. Supplementing this setting with controlled strain is an important future direction. We discussed a possible experimental realization of a pseudo-magnetic field in a  strained membrane that would allow one to probe the predicted edge states. We  demonstrated that in our time-reversal symmetric PLLs  AC modulation of the pseudo-field results in charge pumping. This charge pumping can flow from the superconductor to the normal region in a non-dissipative way as long as the interface is gapped. When the low energy states are present, however, as can be controlled by the Zeeman field, dissipation will occur.

Finally, while we have studied noninteracting electrons, the interplay of interactions with the weakly dispersing PLLs and superconducting correlations may lead to numerous interaction instabilities~\cite{herbut2008pseudomagnetic,Roy2014,Kauppila2016,Xu2018,Peltonen2020} and possibly to realize exotic fractional phases~\cite{Ghaemi2012}.


\vspace{0.1in}
\section{Acknowledgements} E.S and K. S. were supported by the US-Israel Binational
Science Foundation (Grant No. 2016255). E. S. acknowledges support from ARO (W911NF-20-1-0013) and the Israel Science Foundation grant number 154/19. D.S. is supported by the Israel Science Foundation grant No. 1790/18. M.B. is supported by the National Science Foundation grant DMR-2004801. The authors would like to thank Roni Ilan, Moshe Ben Shalom, Felix von Oppen and Eyal Cornfeld for useful discussions.

\bibliographystyle{apsrev4-1}
\bibliography{allref2}

\clearpage

\appendix

\section{Interface dispersion via the Akhmehrov-Beenaker method~\cite{akhmerov2007detection}}
\label{apendix0}

We start with a graphene Hamiltonian containing pseudo-field, intrinsic SOC and Zeeman terms in the valley isotropic basis, as described in Eqs. (\ref{MatrixEquation}), (\ref{ZeemanHamiltonain}) and (\ref{SOCHamiltonain}), and transform it into the BdG equation in Eq.~(\ref{BdGmatrix}).

We wish to describe bound Andreev states at the SN interface, thus we follow the procedure set by  \cite{akhmerov2007detection} and start by solving Eq.~(\ref{BdGmatrix}) in the $x>0$ normal region. The solution follows the canonical lines of PLLs theory in graphene, where we adjust for the presence of SOC and Zeeman terms. We find the eigenvectors
\vspace{-0.1in}
\begin{eqnarray}
\Psi(x,y)=e^{ip_y y}\begin{pmatrix}C_e^{K,\uparrow}\Phi_e^{K,\uparrow}(\xi_+)\\
C_e^{K,\downarrow}\Phi_e^{K,\downarrow}(\xi_+)\\
C_e^{K',\uparrow}\Phi_e^{K',\uparrow}(\xi_-)\\
C_e^{K',\downarrow}\Phi_e^{K',\downarrow}(\xi_-)\\
C_h^{K',\downarrow}\Phi_h^{K',\downarrow}(\xi_+)\\
C_h^{K',\uparrow}\Phi_h^{K',\uparrow}(\xi_+)\\
C_h^{K,\downarrow}\Phi_h^{K,\downarrow}(\xi_-)\\
C_h^{K,\uparrow}\Phi_h^{K,\uparrow}(\xi_-)
\end{pmatrix}\label{eq:FullWaveFunction},
\end{eqnarray}
with the spinors $\Phi$ defined as
\begin{widetext}
\begin{eqnarray}
&&\Phi_e^{K,s_z}(\xi_+)=e^{-\frac{\xi_{+}^{2}}{2}}\begin{pmatrix}-i\left(\mu+\varepsilon+\lambda_I s_z - B_Z s_z\right)H_{\frac{\left(\mu+\varepsilon\right)^{2}-\lambda_I^{2}-2 B_Z\left(\mu+\varepsilon\right)s_z+ B_Z^{2}}{2}-1}(\xi_{+})\\
H_{\frac{\left(\mu+\varepsilon\right)^{2}-\lambda_I^{2}-2 B_Z\left(\mu+\varepsilon\right)s_z+ B_Z^{2}}{2}}(\xi_{+})
\end{pmatrix},
\\
&&\Phi_e^{K',s_z}(\xi_-)=e^{-\frac{\xi_{-}^{2}}{2}}\begin{pmatrix}H_{\frac{\left(\mu+\varepsilon\right)^{2}-\lambda_I^{2}-2 B_Z\left(\mu+\varepsilon\right)s_z+ B_Z^{2}}{2}}(\xi_{-})\\
-i\left(\mu+\varepsilon-\lambda_I s_z- B_Z s_z\right)H_{\frac{\left(\mu+\varepsilon\right)^{2}-\lambda_I^{2}-2 B_Z\left(\mu+\varepsilon\right)+ B_Z^{2}}{2}-1}(\xi_{-})
\end{pmatrix},
\\
&&\Phi_h^{K',s_z}(\xi_+)=e^{-\frac{\xi_{+}^{2}}{2}}\begin{pmatrix}-i\left(\mu-\varepsilon-\lambda_I s_z- B_Z s_z\right)H_{\frac{\left(\mu-\varepsilon\right)^{2}-\lambda_I^{2}-2 B_Z\left(\mu-\varepsilon\right)s_z+ B_Z^{2}}{2}-1}\xi_{+})\\
H_{\frac{\left(\mu-\varepsilon\right)^{2}-\lambda_I^{2}-2 B_Z\left(\mu-\varepsilon\right)s_z+ B_Z^{2}}{2}}(\xi_{+})
\end{pmatrix},
\\
&&\Phi_h^{K,s_z}(\xi_-)=e^{-\frac{\xi_{-}^{2}}{2}}\begin{pmatrix}H_{\frac{\left(\mu-\varepsilon\right)^{2}-\lambda_I^{2}-2 B_Z\left(\mu-\varepsilon\right)s_z+ B_Z^{2}}{2}}(\xi_{-})\\
-i\left(\left(\mu-\varepsilon\right)+\lambda_I s_z- B_Z s_z\right)H_{\frac{\left(\mu-\varepsilon\right)^{2}-\lambda_I^{2}-2 B_Z\left(\mu-\varepsilon\right)s_z+ B_Z^{2}}{2}-1}(\xi_{-})
\end{pmatrix},\label{eq:PhiTerms}
\end{eqnarray}
\end{widetext}
where $s_z=\uparrow\downarrow=\pm1$.
The bound Andreev states on the interface at $x=0$ are described using a boundary condition equation
\begin{eqnarray}
\left(M_{NS}-1\right)\Psi=0 ,\label{eq:BC_equation}
\end{eqnarray} 
where the matrix $M_{NS}$ is given by \cite{titov2006josephson,akhmerov2007detection}
\begin{eqnarray}
M_{NS}=\tau_{0}s_{0}\left(\frac{\varepsilon}{\Delta}-i\sigma_{x}\sqrt{1-\frac{\varepsilon^{2}}{\Delta^{2}}}\right).\label{eq:M_BC_equation}
\end{eqnarray}
This is similar to the BC equation presented in \cite{akhmerov2007detection}, where we include the constraint for transitions between electrons and holes to involve opposite spins via $s_0$ in Eq.~(\ref{eq:M_BC_equation}). Solving Eq.~(\ref{eq:BC_equation}) for the wave function in Eqs.~(\ref{eq:FullWaveFunction})-(\ref{eq:PhiTerms}) at $x=0$, gives the dispersion relation $\varepsilon(p_y)$. By defining 
\begin{align}
&f_{\alpha\pm\lambda_I \mp B_Z }\left(p_y\right)\equiv\frac{H_{\frac{\alpha^{2}-\lambda_I^{2}\mp 2B_Z\alpha+B_Z^{2}}{2}}(p_y)}{\left(\alpha\pm\lambda_I \mp B_Z\right)H_{\frac{\alpha^{2}-\lambda_I^{2}\mp2B_Z\alpha+B_Z^{2}}{2}-1}(p_y)} ,\\
&f_{\alpha\pm\lambda_I\pm B_Z}\left(p_y\right)\equiv\frac{H_{\frac{\alpha^{2}-\lambda_I^{2}\pm2B_Z\alpha+B_Z^{2}}{2}}(p_y)}{\left(\alpha\pm\lambda_I\pm B_Z\right)H_{\frac{\alpha^{2}-\lambda_I^{2}\pm2B_Z\alpha+B_Z^{2}}{2}-1}(p_y)} ,
\end{align}
the four solutions can be written compactly as
\begin{widetext}
\begin{align}
\label{AB1}
&f_{\mu-\varepsilon+s_
{z}\lambda_I+s_{z}B_Z}\left(p_y\right)-f_{\mu+\varepsilon+s_{z}\lambda_I-s_{z}B_Z}\left(p_y\right)=\frac{\sqrt{\Delta^{2}-\varepsilon^{2}}}{\varepsilon}\left(1+f_{\mu+\varepsilon+s_{z}\lambda_I-s_{z}B_Z}\left(p_y\right)f_{\mu-\varepsilon+s_{z}\lambda_I+s_{z}B_Z}\left(p_y\right)\right),\\
\label{AB2}
&f_{\mu-\varepsilon-s_{z}\lambda_I +s_{z}B_Z}\left(-p_y\right)-f_{\mu+\varepsilon-s_{z}\lambda_I -s_{z}B_Z}\left(-p_y\right)=\frac{\sqrt{\Delta^{2}-\varepsilon^{2}}}{\varepsilon}\left(1+f_{\mu+\varepsilon-s_{z}\lambda_I -s_{z}B_Z}\left(-p_y\right)f_{\mu-\varepsilon-s_{z}\lambda_I +s_{z}B_Z}\left(-p_y\right)\right) .
\end{align}
\end{widetext}
We solve Eqs.~(\ref{AB1})-(\ref{AB2}) for $\varepsilon$ as a function of $p_y$ numerically for the case where the Zeeman field $B_Z$ is smaller then the first non-zero PLL to get the spectra plotted in Fig.~\ref{fig:res}. 



\begin{figure}[t]
\includegraphics[width=0.9\columnwidth, keepaspectratio=true]{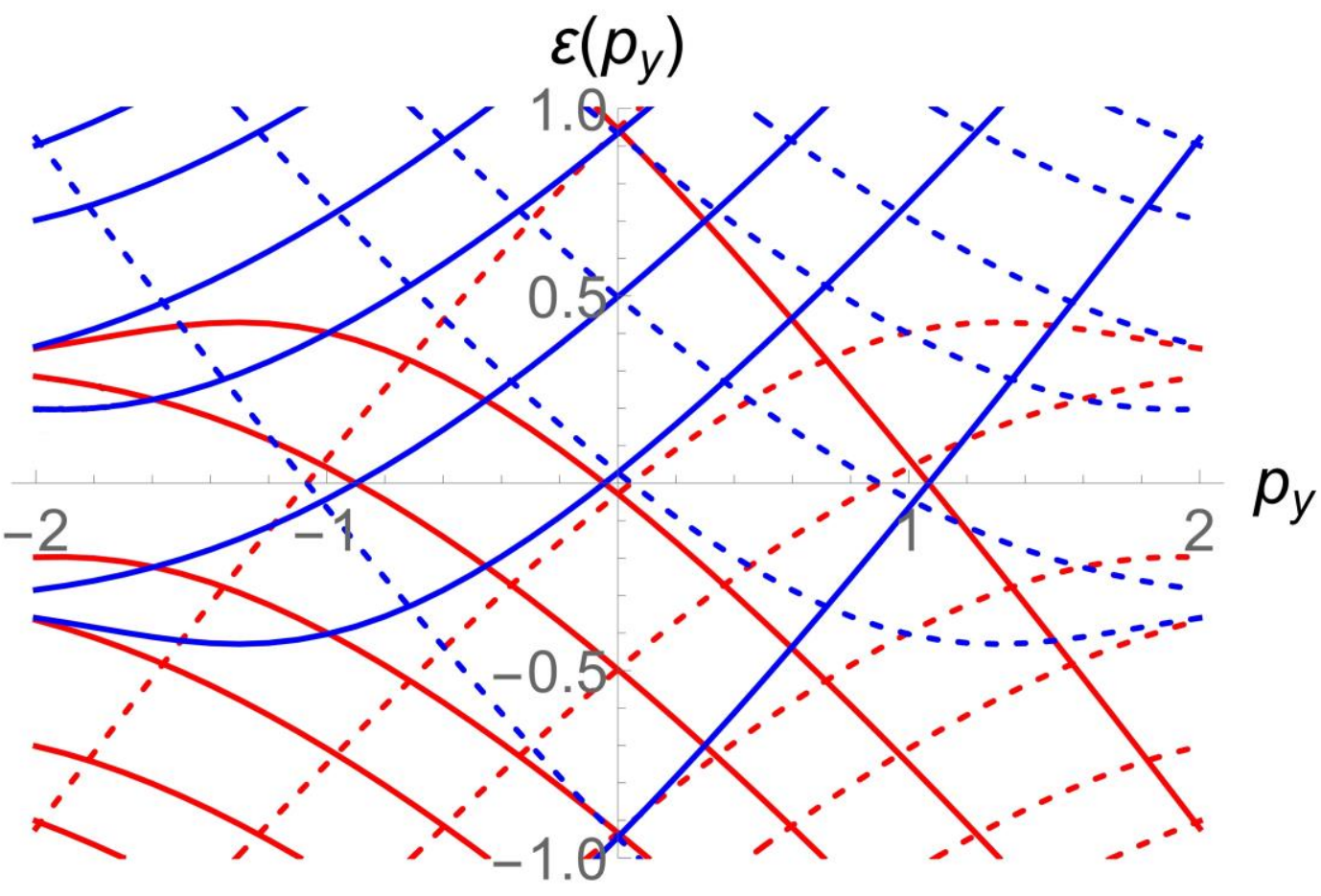}\hfill
\caption{Dispersion relation $\varepsilon(p_y)$ as in Fig.~\ref{fig:res}, with the parameters $\Delta = 10$, $\mu=0$ and $B_Z=\sqrt{3}$, leading to a total of six crossings.}
\label{fig:NextLL}
\end{figure}

As $B_Z$ increases beyond the energy of the $n\ge1$ PLLs, more zero energy crossings are created. To simplify the picture, we focus on $\mu=0$, and plot in Fig.~\ref{fig:NextLL} the spectrum for a value of $B_Z$ between the first and second PLLs, $B_Z=\sqrt{3}$. When $B_Z$ surpasses the energy of each $n>0$ PLL, we see eight new bands that go through $E=0$, leading to four new crossings. Two crossings appear when $B_Z$ is equal to the energy of the LL, and another two appear when $B_Z$ is larger.

\section{Interface dispersion via tight-binding calculations}
\label{apendix2}

\begin{figure}[t]
\includegraphics[width=0.89\columnwidth, keepaspectratio=true]{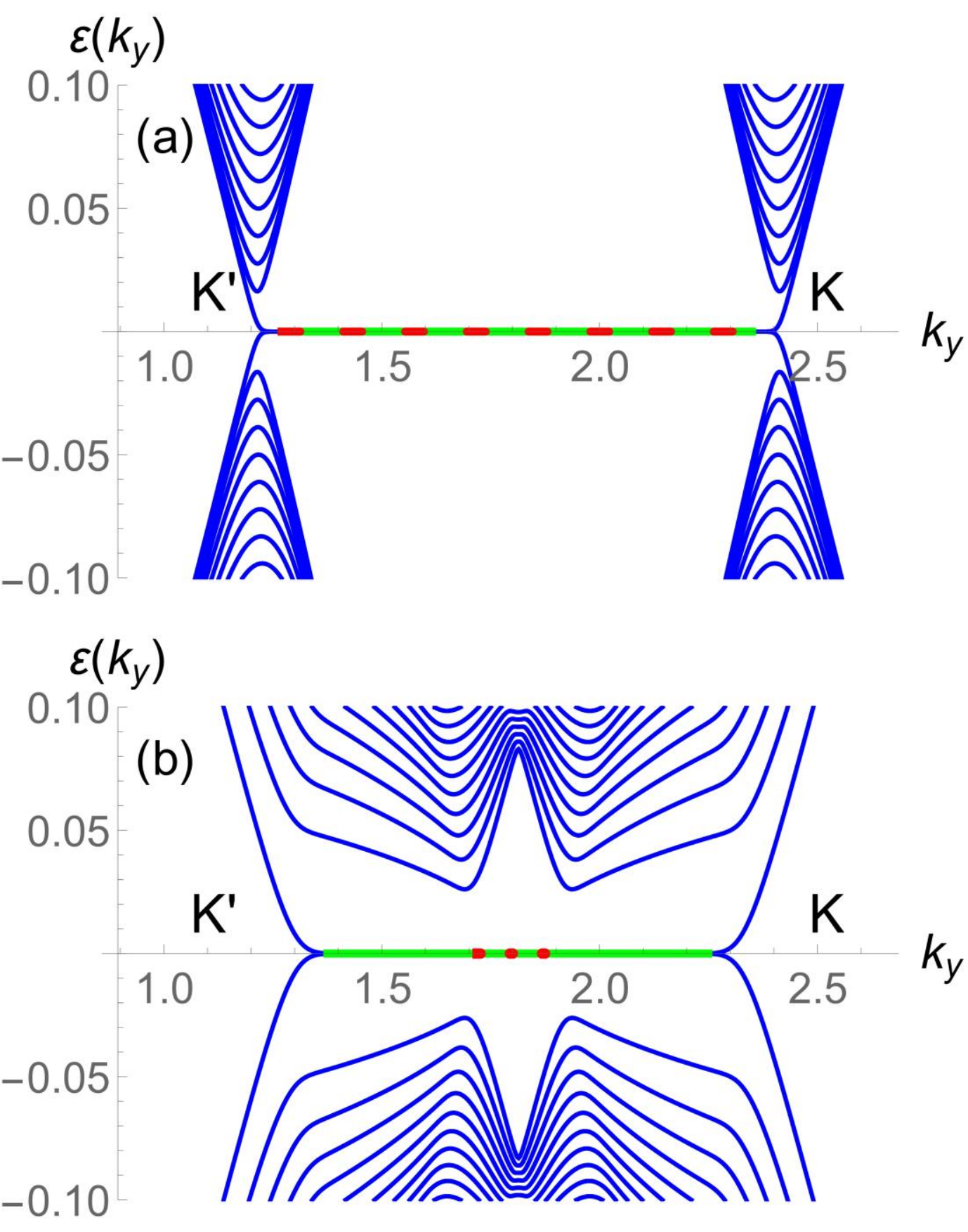}\hfill\\
\caption{Spectrum of  a graphene ribbon described by the tight-binding model in Eq.~(\ref{eq:TBGrapheneTOT}) without (a) and with (b) a pseudo field generated by $\delta t=0.006$. The system size is $N_n=140$.  The green and red segments represent the zigzag edge states on the bottom and top of the ribbons, respectively, whose inversion symmetry is broken by the strain gradient in (b).}   
\label{fig:TBN}\vspace{-0.1in}
\end{figure}

\begin{figure}[t]
\includegraphics[width=0.6\columnwidth, keepaspectratio=true]{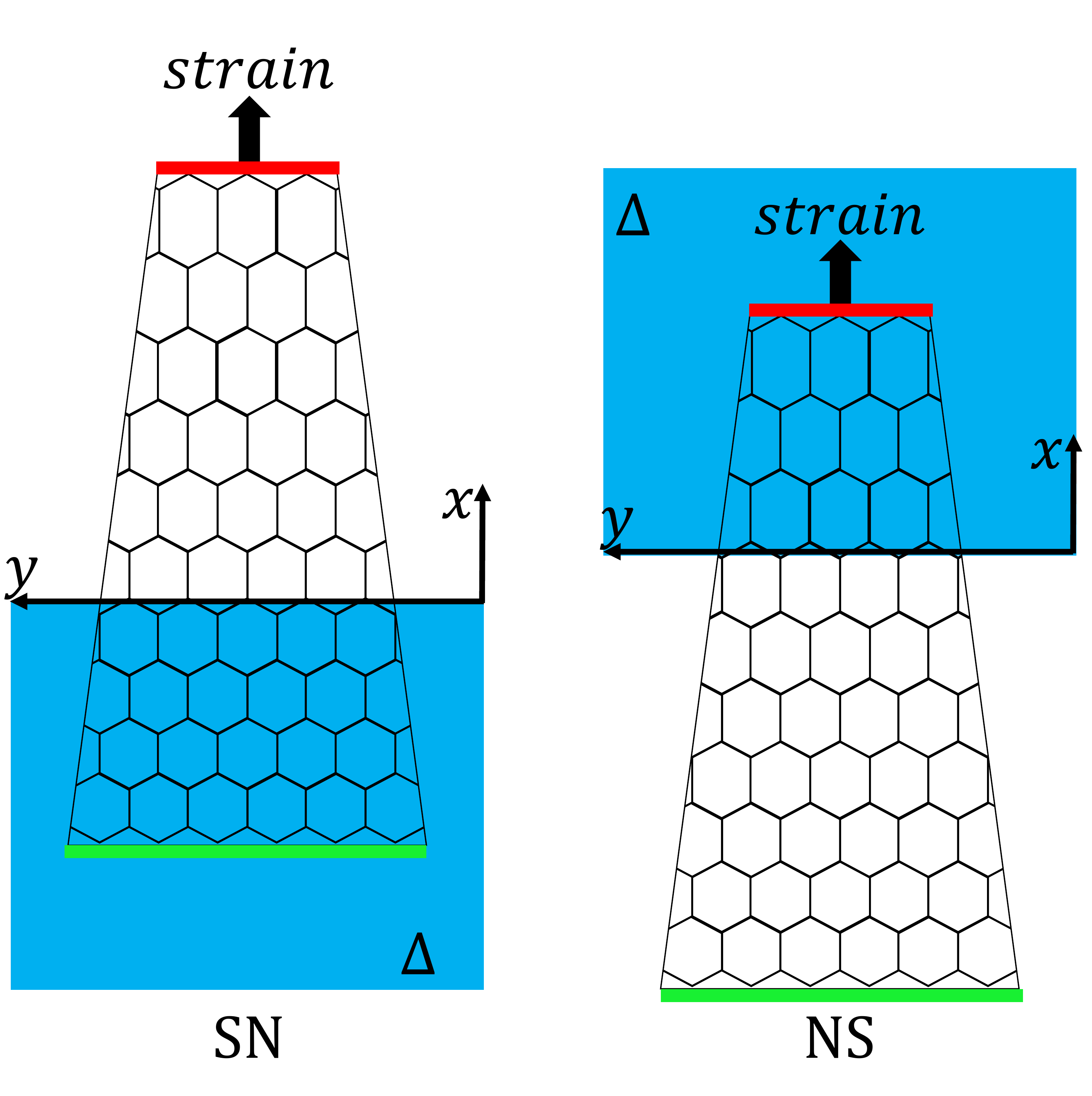}\hfill
\caption{Illustration of the two types of normal-superconducting interfaces we investigate in Fig.~\ref{fig:TBAll}. In the SN configuration the superconducting region is at the bottom of the graphene ribbon, and in the NS configuration its at the top. The zigzag edge states are color coded in accordance with the zigzag bands in the spectra in Figs.~\ref{fig:TBN} and \ref{fig:TBAll}.}   
\label{fig:SNNS}\vspace{-0.0in}
\end{figure}

\begin{figure*}[t]
\includegraphics[width=0.9\textwidth, keepaspectratio=true]{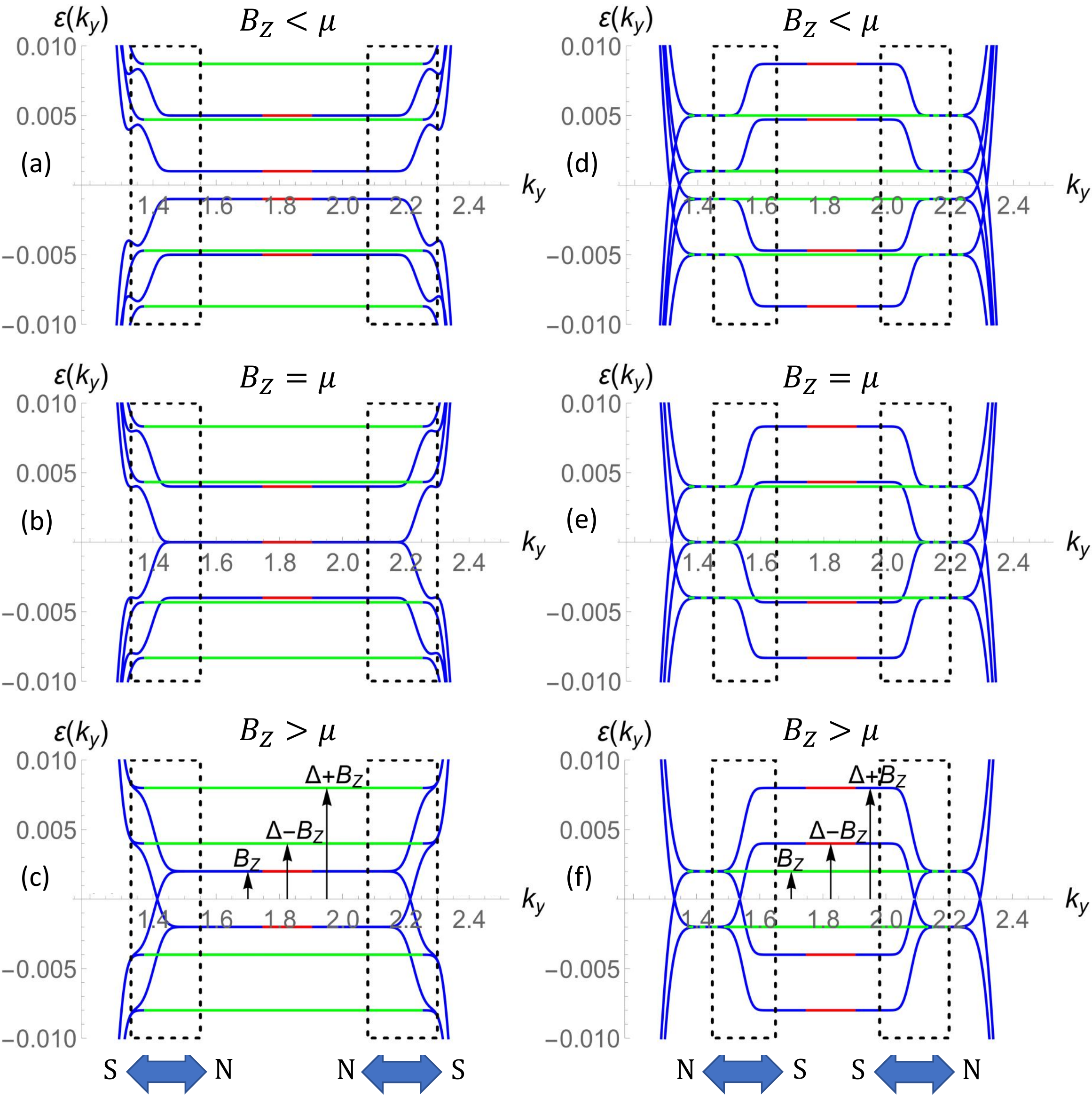}\hfill
\caption{Spectra of an SN [(a)-(c)] and NS [(d)-(c)]  graphene ribbon interface described by the tight-binding model in Eq.~(\ref{eq:TBGraphene}), with pseudo field, a Zeeman field described by Eq.~(\ref{eq:TBZeeman}) and a superconducting potential described by Eq.~(\ref{eq:TBSC}). The size of the normal region is $N_n=90$, the size of the superconducting region is $N_s=50$, and we use the parameters $\delta t=0.06$, $\Delta=0.006$ and $B_Z=0.002$. In (a) and (d) $\mu=0.3$ and we see a gap on the SN interface, in (b) and (e) $\mu=0.2$ and this gap closes, and in (c) and (f) $\mu=0.1$ and we observe the helical states described in the main text. The dashed rectangles mark the interface between the normal and superconducting regions around each node. The green and red lines are the zigzag edge states on the bottom and top of the ribbons, respectively.}
\label{fig:TBAll}\vspace{-0.2in}
\end{figure*}

To corroborate the low energy calculations, we construct a tight-binding model of the system presented in the main text. We look at a graphene hexagonal lattice in a ribbon geometry, translationally invariant in the $\hat{y}$ (zigzag) direction and finite in the $\hat{x}$ (armchair) direction, where the directions are described in Fig. \ref{fig1} in the main text. To create the pseudo field, we linearly change the hopping amplitude in the $\hat{x}$ direction \cite{Ghaemi2012,WenYu2013,lee2017fractional,lantagne2020dispersive}. 
The full tight-binding Hamiltonian is
\be
\label{eq:TBGrapheneTOT}
H_{tb}=H_{G}+H_{Z}+H_{SC}.
\ee
Here, the graphene ribbon with pseudo field Hamiltonian reads   
\begin{eqnarray}
  \label{eq:TBGraphene}
  H_{G}&&=\sum_{s}\int\frac{dk_{y}}{2\pi}\left[-\sum_{\sigma}\sum_{i=1}^{N}\mu c_{i,\sigma,s}^{\dagger}\left(k_{y}\right)c_{i,\sigma,s}\left(k_{y}\right)\right.\\
  &&-\sum_{i=1}^{N}t\cos\left(\frac{\sqrt{3}}{2}ak_{y}\right)c_{i,B,s}^{\dagger}\left(k_{y}\right)c_{i,A,s}\left(k_{y}\right)\nonumber\\
  &&\left.-\sum_{i=2}^{N}\left(t-\delta t\left(i\right)\right)\left(c_{i-1,B,s}^{\dagger}\left(k_{y}\right)c_{i,A,s}\left(k_{y}\right)\right)+h.c\right],\nonumber
\end{eqnarray}
where $s$ runs over spin, $i$ runs over the discretized layers of the lattice in the $\hat{x}$ direction, $t$ is the hopping amplitude between nearest neighbours, $\delta t(i)$ is the change in the hopping amplitude for hopping between adjacent layers, which is a function of the layer index $i$, and $A/B$ is the honeycomb sublattice index. In what follows we set $a=1$, $t=1$ and $dt=0.006$. We plot the spectrum with and without a pseudo-field in Fig.~\ref{fig:TBN}.



We add a Zeeman field 
\begin{align}
H_{Z}=\sum_{\lambda,i}\int\frac{dk_{y}}{2\pi}B_Z\left[c_{i,\lambda,\uparrow}^{\dagger}\left(k_{y}\right)c_{i,\lambda,\uparrow}\left(k_{y}\right)-\left(\uparrow\rightarrow\downarrow\right)\right],
\label{eq:TBZeeman}
\end{align}
where $\lambda=A,B$ is a sublattice index, and $B_Z$ is the Zeeman amplitude. Finally, we induce superconductivity in the regions specified below, with the following Hamiltonian \cite{lee2017fractional}

\begin{eqnarray}
\label{eq:TBSC}
H_{SC}&=&\sum_{\lambda,i \in SC}\int\frac{dk_{y}}{2\pi}\left[\Delta c_{i,\lambda,\uparrow}^{\dagger}\left(k_{y}\right)c_{i,\lambda,\downarrow}^{\dagger}\left(-k_{y}\right)\right.\nonumber\\
&&\left.+h.c-\left(\uparrow\rightleftarrows\downarrow\right)\right] ,
\end{eqnarray}
where $\Delta$ is the superconductive potential. Here $i \in SC$ refers to the sites that belong to the proximitized region.

We compute the spectrum of the tight-binding model for an interface between a normal and a superconducting region. We denote $N_s$ the size of the superconductive region ($\Delta>0$), and $N_n$ the size of the normal region ($\Delta=0$). Unlike in the Akhmerov-Beenakker method and the low energy calculations, the ribbon geometry produces zigzag edge states in the spectrum, which are different on each of the edges due to the pseudo field. We thus look at both possible configurations for the interface, i.e. one where the superconducting region is at the top edge of the ribbon (NS), and the other is where the superconductive region at the bottom edge of the ribbon (SN). Both configurations are plotted in Fig.~\ref{fig:SNNS}. The strain gradient breaks the inversion symmetry making these configurations nonequivalent.

\begin{figure}[t]
\includegraphics[width=0.89\columnwidth, keepaspectratio=true]{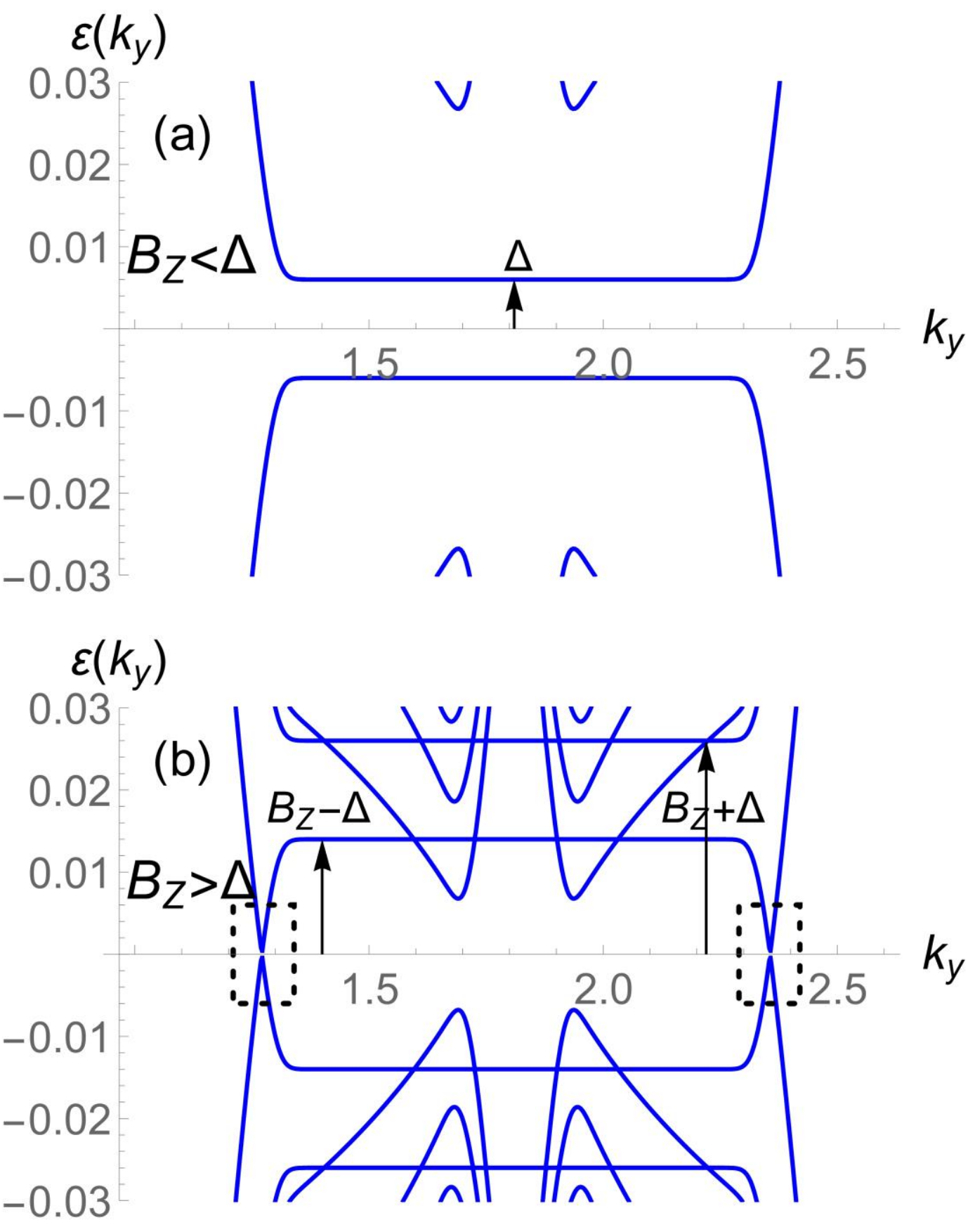}\hfill
\caption{Spectrum of fully superconducting system with pseudo field. We set $\Delta=0.006$, $\delta t=0.006$ and $\mu=0$. In (a) $B_Z=0<\Delta$ and in (b)$B_Z=0.008>\Delta$. The system size is $N_s=140$.}
\label{fig:TBNSY}\vspace{-0.2in}
\end{figure}

\begin{figure}[b]
\includegraphics[width=0.89\columnwidth, keepaspectratio=true]{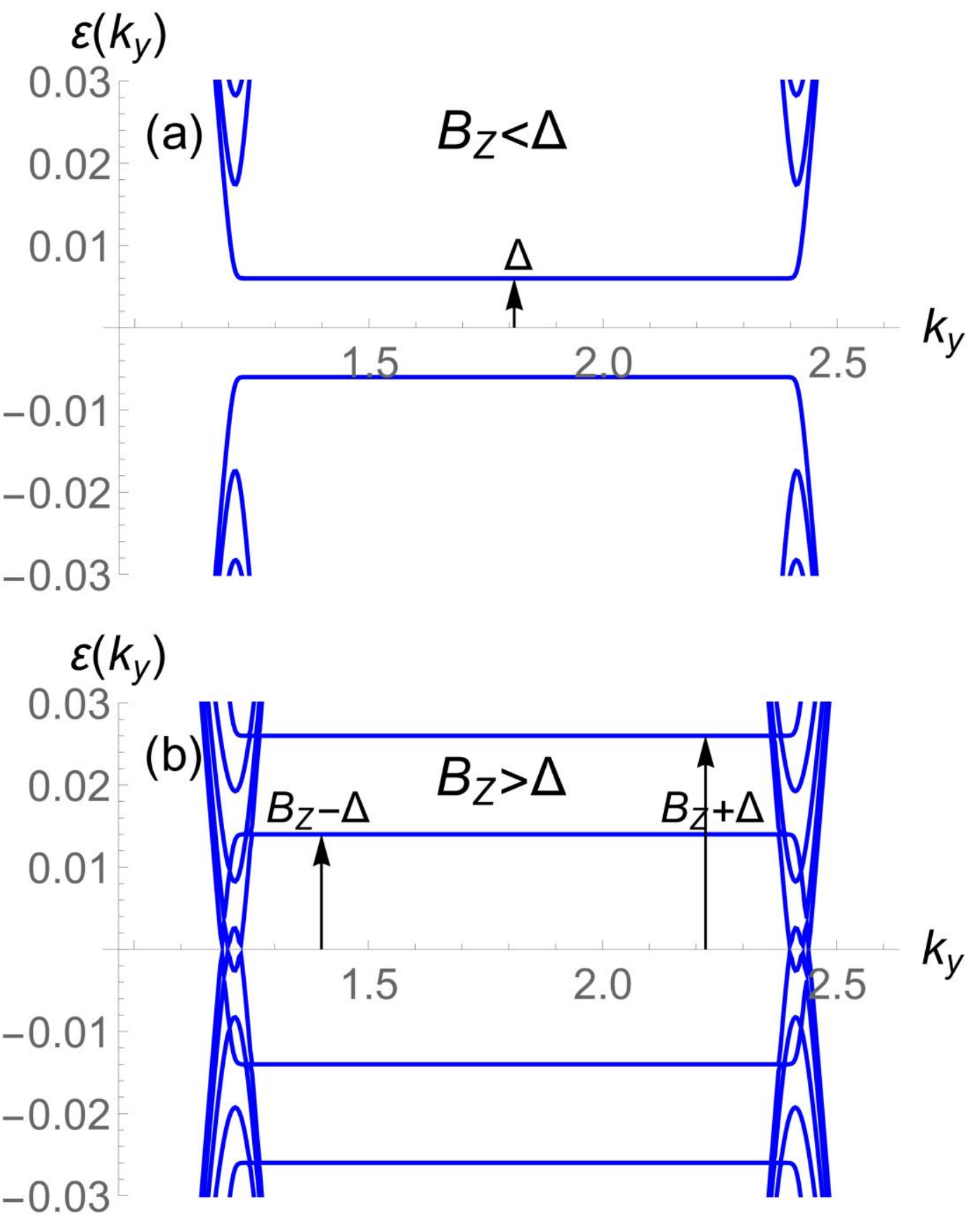}\hfill
\caption{Same as in Fig.~\ref{fig:TBNSY}, with $\delta t=0$.}
\label{fig:TBNSX}\vspace{-0.2in}
\end{figure}

We plot the spectra of the three scenarios discussed in the main text: $B_Z<\mu$ , $B_Z=\mu$ and $B_Z>\mu$ for the SN and NS geometries in Fig.~\ref{fig:TBAll}.  The zigzag edge states on the bottom of the ribbon are colored green, and the zigzag edge states on the top of the ribbon are colored red.  The position-momentum coupling is such that around the K node increasing momentum corresponds to decreasing $x-$position, while around the K' node increasing momentum corresponds to increasing position. As seen in Fig.~\ref{fig:TBAll}, specifically in panels (c) and (f) of the mentioned figures, the tight-binding calculations reproduces the helical states when $B_Z>\mu$.

\subsection{Superconducting-vacuum interface}
\label{apendix2_1}

We now consider the zigzag graphene ribbon under (i) uniform Zeeman splitting $B_Z$, (ii) uniform pairing
$\Delta$, and (iii) strain gradient leading to PLLs. The resulting tight-binding spectrum is shown in Fig.~\ref{fig:TBNSY}. We see that for $B_Z > \Delta$ the interface with vacuum supports gapless edge states. Therefore, our class $D$ superconductor supports a  topological nontrivial phase.

As discussed in Sec.~\ref{se:AZ}, due to the $S^z$ conservation our system maps to two models classified by the Altland-Zirnbauer class $D$ superconductor with $\mathbb{Z}$ index in each sector. This double-copy allows to accommodate the presence of counter propagating modes at each interface, as seen in the small boxes in Fig.~\ref{fig:TBNSY}(b) (similar to a system of spin-up electrons under a positive orbital magnetic field and spin-down electrons under opposite field). As we discussed in depth above, the nature of each chiral mode is that of a Dirac fermion, which can be decomposed into two Majorana fermions. Hence in each sector our topological superconductor exhibits an even $\mathbb{Z}$ index. Presumably this can be associated with the fact that pairing connects two disconnected Fermi points in our system.

Turning off the strain, the spectrum is shown in Fig.~\ref{fig:TBNSX}. We can see that similar to the case with PLLs, the spectrum becomes gappless for $B_Z > \Delta$ even in the absence of PLLs. However, the present state is gapless in the bulk. Indeed the entire Dirac cones, which have quantized levels due to the finite stripe, cross the $\varepsilon=0$ line.  On the other hand in the strained case of Fig.~\ref{fig:TBNSY} the gapped modes are localized at the boundary.

\end{document}